# AUTOMATIC ONTOLOGY GENERATION FOR DATA MINING USING FCA AND CLUSTERING

Amel Grissa Touzi[1], Hela Ben Massoud[1] and Alaya Ayadi[1]

[1]*Ecole Nationale d'Ingénieurs de Tunis, Tunisia*
[2]*Department of Technologies of Information and Communications*
*amel.touzi@enit.rnu.tn, benmessaoudhella@gmail.com, alaa.ayadi@gmail.com*

Abstract: Motivated by the increased need for formalized representations of the domain of Data Mining, the success of using Formal Concept Analysis (FCA) and Ontology in several Computer Science fields, we present in this paper a new approach for automatic generation of Fuzzy Ontology of Data Mining (*FODM*), through the fusion of conceptual clustering, fuzzy logic, and FCA. In our approach, we propose to generate ontology taking in consideration another degree of granularity into the process of generation. Indeed, we suggest to define an ontology between classes resulting from a preliminary classification on the data. We prove that this approach optimize the definition of the ontology, offered a better interpretation of the data and optimized both the space memory and the execution time for exploiting this data.

## 1 INTRODUCTION

While knowledge discovery in databases (KDD) and Data Mining have enjoyed great popularity and success in the recent years, there is have been a distinct lack of a generally accepted framework that would describe and unify the area of data mining. The present lack of such a framework is perceived as an obstacle to the further development of the field.

In (Yang, 2006), Yang and Wu collected the opinions of a number of outstanding data mining researchers about the most challenging problems in data mining research. Among the ten topics considered as the most important ones and worthy of further research, the development of an unifying framework for data mining is listed first.

Many researchers in the field of data mining have tried to construct ontology for data mining targeted to solve some specific problems. Most of the developments are with the aim of automatic planning of data mining workflows (Bernstein, 2005, Zakova, 2008; Kalousis, 2008). Some of them are concerned with the description of the data mining services on the grid (Brezany, 2007; Cannataro, 2003). Other different directions about possible interactions among ontology and Formal Concept Analysis (FCA) aim at modeling concepts are being explored in different field like the semantic web (Tho, 2006) and the text documents (Cimiano, 2004).

The problem of these ontology is that they are not constructed to describe the complete domain of data mining but are simply made with a specific task in mind. Almost all the proposed ontology, with the small exception of the work presented in (Zakova, 2008), deal with propositional data mining algorithms and do not take into account the existence of data mining algorithms for mining structured data. Moreover, all the approaches are superficial in sense that they look at data mining algorithms as black boxes, describing them only by their inputs and outputs, not trying to describe the basic components of the algorithms.

In our point of view, the limits of these approaches consist in extracting this ontology departing from the data or a data variety, which may be huge. To cure all these problems, we propose a new approach for generation of the ontology takes in consideration another degree of granularity into the process of this generation. Indeed, we propose to define ontology between classes resulting from a preliminary classification on the data. The data classification is to divide a data set into subsets, called classes, so that all data in the same class are

similar and data from different classes are dissimilar. Thus:
- The number of clusters generated by a classification algorithm is always less than the number of objects starting on which we apply the classification algorithm.
- All objects belonging to the same cluster have the same properties.

We prove that this approach optimize the definition of the ontology, offered a better interpretation of the data and optimized both the space memory and the execution time for exploiting this data.

The rest of the paper is organized as follows: section 2 presents the basic concepts of ontology and Formal Concept Analysis (FCA). Section 3 presents related work; Section 4 presents our motivation for this work. Section 5 describes our new approach for the automatic generation of Fuzzy Ontology of Data Mining, called FODM. Section 6 enumerates the advantages and validates the proposed approach. We finish this paper with a conclusion and a presentation of some future works.

## 2 BASIC CONCEPTS

### 2.1 Ontologies

Ontologies (Chandrasekaran,1999) are content theories about the classes of individuals, properties of individuals, and relations between individuals that are possible in a specified domain of knowledge. They define the terms for describing our knowledge about the domain. An ontology of a domain is beneficial in establishing a common (controlled) vocabulary for the describing the domain of interest. This is important for unification and sharing of knowledge about the domain and connecting with other domains.

In reality, there is no common formal definition of what ontology is. However, most approaches share a few core items such as : concepts, a hierarchical IS-A-relation, and further relations. For the sake of generality, we do not discuss more specific features like constraints, functions, or axioms in this paper, instead we formalize the core in the following way:

**Definition:** A *(core) ontology* is a tuple
$O = (C, is\_a, R, \sigma)$ where
- C is a set whose elements are called concepts,
- is_a is a partial order on C  C (i. e., a a binary relation is_a $\subseteq$ C X C which is reflexive, transitive, and anti symmetric),
- R is a set whose elements are called relation names (or relations for short),
- $\sigma : R \rightarrow C^+$ is a function which assigns to each relation name its arity.

In the last years, several languages have been developed to describe ontologies. As example, we can cite, the Resource Description Framework (RDF) (Lassila, 1999; Klyne, 2004), the Ontology Web Language (OWL) (Bechhofer, 2004) and extension of OWL language like OWL 2 (Cuenca-Grau, 2008) or Fuzzy OWL(Bobillo, 2010).

Also, the number of environments and tools for building ontologies has grown exponentially. These tools are aimed at providing support for the ontology development process and for the subsequent ontology usage. Among these tools we can mention most relevant: Ontolinguav (Farquhar, 1996), WebOnto (Domingue, 1999), WebODE(Arpirez, 2001), Protégé-2000 ( Noy, 2000), OntoEdit(Sure, 2002) and OilEd(Bechhofer, 2001).

### 2.2 Fuzzy Conceptual Scaling and FCA

Formal concept analysis (FCA) is a method for data analysis, knowledge representation and information management. It was proposed by Rudolf Wille in 1982 (Wille, 1982). In recent years, FCA has grown into an international research community with applications in many disciplines, such as linguistics, software engineering, psychology, medicine, AI, database, library science, ecology, information retrieval, ontology construction , etc.

FCA starts with the notion of a formal context specifying which objects have what attributes and thus a formal context may be viewed as a binary relation between the object set and the attribute set with the values of 0 and 1. In (Quan, 2004), an ordered lattice extension theory has been proposed: Fuzzy Formal Concept Analysis (FFCA), in which uncertainty information is directly represented by a real number of membership value in the range of [0,1]. This number is equal to the similarity defined as follow:

**Definition.** The similarity of a fuzzy formal concept $C_1 = (\varphi(A_1), B_1)$ and its subconcept $C_2 = (\varphi(A_2), B_2)$ is defined as:

$$S(C_1, C_2) = \frac{|\varphi(A_1) \cap \varphi(A_2)|}{|\varphi(A_1) \cup \varphi(A_2)|}$$

where $\cap$ and $\cup$ refer intersection and union operators on fuzzy sets, respectively (Zadeh, 1975).

In (Grissa, 2009), we showed as these FFCA are very powerful as well in the interpretation of the results of the Fuzzy Clustering and in optimization of the flexible query.

## 3 RELATED WORK

Usually the ontology building is performed manually, but researchers try to build ontology automatically or semi automatically to save the time and the efforts of building the ontology. We survey in this section the most important approaches that generate ontologies from data.

Clerkin et al. used concept clustering algorithm (COBWEB) to discover automatically and generate ontology. They argued that such an approach is highly appropriate to domains where no expert knowledge exists, and they propose how they might employ software agents to collaborate, in the place of human beings, on the construction of shared ontologies (Clerkin, 2001).

Blaschke et al. presented a methodology that creates structured knowledge for gene-product function directly from the literature. They apply an iterative statistical information extraction method combined with the nearest neighbour clustering to create ontology structure (Blaschke, 2002).

Formal Concept Analysis (FCA) is an effective technique that can formally abstract data as conceptual structures (Ganter, 2005). Quan et al. proposed to incorporate fuzzy logic into FCA to enable FCA to deal with uncertainty in data and interpret the concept hierarchy reasonably, the proposed framework is known as Fuzzy Formal Concept Analysis (FFCA).They use FFCA for automatic generation of ontology for scholarly semantic web (Quan, 2004).

Dahab et al. presented a framework for constructing ontology from natural English text namely TextOntEx. TextOntEx constructs ontology from natural domain text using semantic pattern-based approach, and analyzes natural domain text to extract candidate relations, then maps them into meaning representation to facilitate ontology representation (Dahab, 2007)

Wuermli et al. used different ways to build ontologies automatically, based on data mining outputs represented by rule sets or decision trees. They used the semantic web languages, RDF, RDF-S and DAML+OIL for defining ontologies (Wuermli 2003).

## 4 MOTIVATION

The motivation for developing an ontology of data mining is multi-fold.
- The area of data mining is developing rapidly and one of the most challenging problems deals with developing a general framework for data mining. By developing an ontology of data mining, we are taking one step towards solving this problem.
- The traditional task of the knowledge engineer is to translate the knowledge of the expert into the knowledge base of the expert system. Knowledge engineer uses ontology to represent the knowledge of the domain expert. Due to of the difficulty to find a domain expert and the needing for updating the knowledge represented in the ontology frequently, our challenge is to define a system for building ontology automatically from the database.
- There exist several proposals for ontology of data mining but all of them are light-weight, aimed at covering a particular use-case in data mining and are of a limited scope and highly use-case dependent. Data mining is a domain that needs a heavy-weight ontology with a broader scope, where much attention is paid to the rigorous meaning of each entity, semantically rigorous relations between entities and compliance to an upper level ontology and the domains of application.

In our point of view, the limits of these approaches consist in extracting this ontology departing from the data or a data variety, which may be huge. To cure all these problems, we propose a new approach for generation of the ontology using conceptual clustering, fuzzy logic, and FCA. Indeed, we propose to define ontology between classes resulting from a preliminary classification on the data. The data classification is to divide a data set into subsets, called classes, so that all data in the same class are similar and data from different classes are dissimilar.

## 5 PRESENTATION OF THE FUZZY ONTOLOGY OF DATA MINING: FODM

### 5.1 Principe of the FODM

In this section, we present the architecture of the Fuzzy Ontology of Data Mining (FODM) approach

and the process for constructing fuzzy ontology. Our FODM approach takes the database records and provides the corresponding fuzzy ontology Figure 1 shows the proposed approach.

In our approach, we propose to generate this ontology taking in consideration another degree of granularity into the process of generation. Indeed, we propose to define the ontology between classes resulting from a preliminary classification on the data.

The FODM approach is organized according to two following principal steps.

1. **Data Organization step:** it permits to organize the database records in homogeneous clusters having common properties. This step gives a certain number of clusters for each attribute. Each tuple has values in the interval [0,1] representing these membership degrees according to the formed clusters. Linguistic labels, which are fuzzy partitions, will be attributed on attribute's domain. This step consists of TAH's and MTAH generation of relieving attributes. This step is very important in the FODM process because it allows us to define and interpreter the distribution of objects in the various clusters.
2. **Fuzzy Ontology Generation step:** The second step consists in on **the construct on Fuzzy Ontology**. It consists in deducing the Fuzzy Cluster Lattice corresponding to MTAH lattice generated in the first step, then to generate **Ontology Extent and Intent Classes, Ontology hierarchical Classes**. **Ontology Relational Classes and Fuzzy Ontology.**

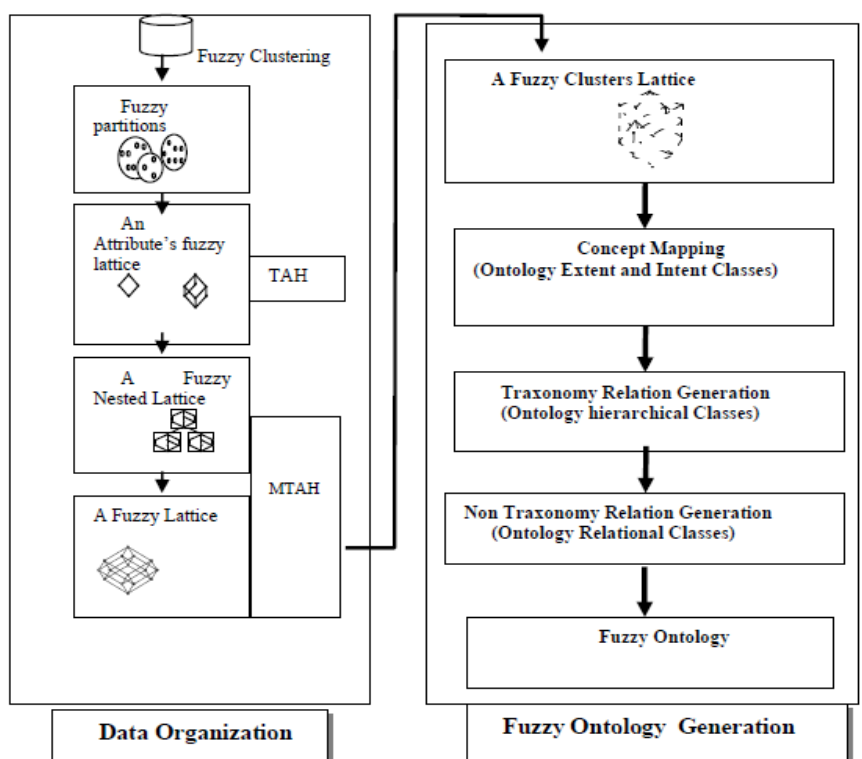

Figure 1: Presentation of the Fuzzy Ontology of Data Mining approach

## 5.2 Theoretical Foundation of the KDD model

In this part, we present the theoretical foundations of the proposed approach, based on the following properties:
**Properties 1**

- The number of clusters generated by a classification algorithm is always lower than the number of starting objects to which one applies the classification algorithm
- All objects belonging to one same cluster have the same proprieties. These characteristics can be deduced easily knowing the center and the distance from the cluster.

- The size of the lattice modeling the properties of the clusters is lower than the size of the lattice modeling the properties of the objects.
- The management of the lattice modeling the properties of the clusters is optimum than the management of the lattice modeling the properties of the objects.

**Properties 2**

Let C1, C2 be two clusters, generated by a classification algorithm and verifying the properties p1 and p2 respectively. Then the following properties are equivalent:

$C1 \Rightarrow C2$ (CR) $\Leftrightarrow$

- $\forall$ object O1 $\in$ C1 => O1 $\in$ C2 (CR)
- $\forall$ object O1 $\in$ C1, O1 checks the property p1 of C1 and the property p2 of C2. (CR)

**Properties 3**

Let C1, C2 and C3 be three clusters generated by a classification algorithm and verifying the properties p1, p2 and p3 respectively. Then the following properties are equivalent:

C1, C2 => C3 (CR)

$\Leftrightarrow$

- $\forall$ object O1 $\in$ C1 $\cap$ C2 => O1 object $\in$ C3 (CR)
- $\forall$ object O1 $\in$ C1 $\cap$ C2 then O1 checks the properties p1, p2 and p3 with (CR)

The proof of the two properties rises owing to the fact that all objects which belong to a same cluster check necessarily the same property as their cluster.

## 5.3 Data Organization Step

This step gives a certain number of clusters for each attribute. Each tuple has values in the interval [0,1] representing these membership degrees according the formed clusters. Linguistic labels, which are fuzzy partitions, will be attributed on attribute's domain. This step consists of TAH's and MTAH generation of relieving attributes. This step is very important in KDD Process because it allows to define and interpreter the distribution of objects in the various clusters.

**Example**: Let a relational database table presented by Table1 containing the list of AGE and SALARY of Employee. Table 2 presents the results of fuzzy clustering (using Fuzzy C-Means (Sun, 2004) ) applied to Age and Salary attributes. For Salary attribute, fuzzy clustering generates three clusters (C1,C2 and C3). For AGE attribute, two clusters have been generated (C4 and C5). In our example, $\alpha - Cut$ (Salary) = 0.3 and $\alpha - Cut$ (Age) = 0.5, so, the Table 2 can be rewriting as show in Table 3.

The corresponding fuzzy concept lattices of fuzzy context presented in Table 3, noted as TAH's are given by the line diagrams presented in the Figure 2 and 3.

Table 1: This A relational database table.

|    | SALARY | AGE |
|----|--------|-----|
| t1 | 800    | 30  |
| t2 | 600    | 35  |
| t3 | 400    | 26  |
| t4 | 900    | 40  |
| t5 | 1000   | 27  |
| t6 | 500    | 30  |

Table 2: This Fuzzy Conceptual Scales for age and salary attributes

|    | SALARY |     |     | AGE |     |
|----|--------|-----|-----|-----|-----|
|    | C1     | C2  | C3  | C4  | C5  |
| t1 | 0.1    | 0.5 | 0.4 | 0.5 | 0.5 |
| t2 | 0.3    | 0.6 | 0.1 | 0.4 | 0.6 |
| t3 | 0.7    | 0.2 | 0.1 | 0.7 | 0.3 |
| t4 | 0.1    | 0.4 | 0.5 | 0.2 | 0.8 |
| t5 | -      | 0.5 | 0.5 | 0.6 | 0.4 |
| t6 | 0.5    | 0.5 | -   | 0.5 | 0.5 |

Table 3: This Fuzzy Conceptual Scales for age and Salary attributes with $\alpha - Cut$.

|    | SALARY |     |     | AGE |     |
|----|--------|-----|-----|-----|-----|
|    | C1     | C2  | C3  | C4  | C5  |
| t1 | -      | 0.5 | 0.4 | 0.5 | 0.5 |
| t2 | 0.3    | 0.6 | -   | -   | 0.6 |
| t3 | 0.7    | -   | -   | 0.7 | -   |
| t4 | -      | 0.4 | 0.5 | -   | 0.8 |
| t5 | -      | 0.5 | 0.5 | 0.6 | -   |
| t6 | 0.5    | 0.5 | -   | 0.5 | 0.5 |

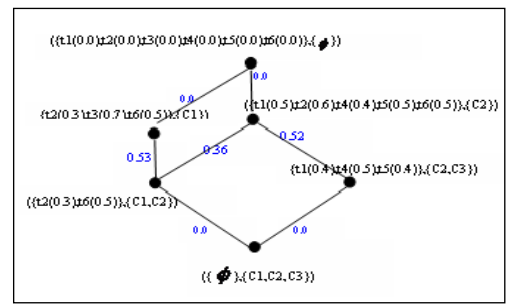

Figure 2: Salary TAH

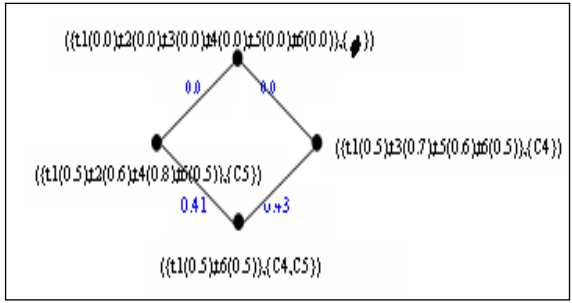

Figure 3: Age TAH

The minimal value (resp. maximal) of each cluster corresponds on the lower (resp. higher) interval terminal of the values of this last. Each cluster of a partition is labeled with a **linguistic labels** provided by the user or a domain expert.

For example, the fuzzy labels *young* and *adult* could belong to a partition built over the domain of the attribute *AGE*. Also, the fuzzy labels *low*, *Medium* and *High*, could belong to a partition built over the domain of the attribute *Salary*. The Table 4 presents the correspondence of the linguistic labels and their designations for the attributes Salary and Age. The corresponding fuzzy concept lattices of fuzzy context presented in Table 5, noted as TAH's are given by the line diagrams presented in Figure 1 and 2.

TABLE 4: THIS CORRESPONDENCE OF THE LINGUISTIC LABELS AND THEIR DESIGNATIONS

| Attribute | Linguistic labels | Designation |
|---|---|---|
| Salary | Low | C1 |
| Salary | Medium | C2 |
| Salary | High | C3 |
| Age | Young | C4 |
| Age | Adult | C5 |

TABLE 5: THIS FUZZY CONCEPTUAL SCALES FOR AGE AND SALARY ATTRIBUTES WITH $\alpha - Cut$.

|  | SALARY | | | AGE | |
|---|---|---|---|---|---|
|  | Low C1 | Medium C2 | High C3 | Young C4 | Adult C5 |
| t1 | - | 0.5 | 0.4 | 0.5 | 0.5 |
| t2 | 0.3 | 0.6 | - | - | 0.6 |
| t3 | 0.7 | - | - | 0.7 | - |
| t4 | - | 0.4 | 0.5 | - | 0.8 |
| t5 | - | 0.5 | 0.5 | 0.6 | - |
| t6 | 0.5 | 0.5 | - | 0.5 | 0.5 |

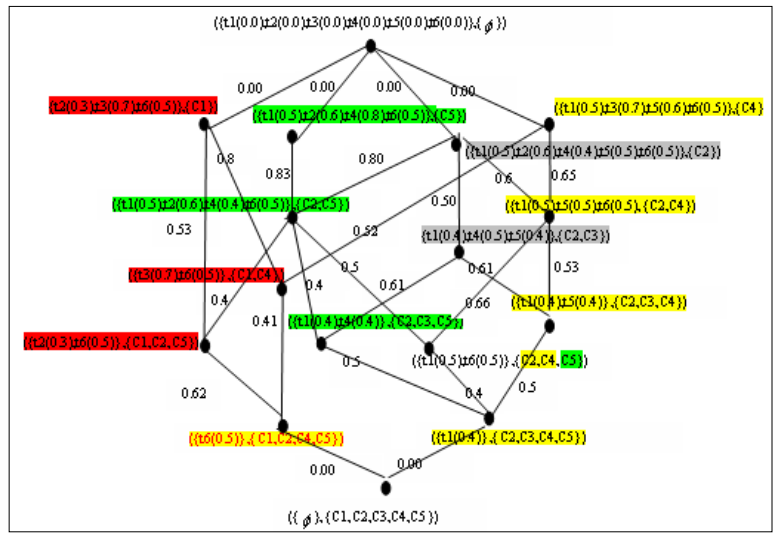

Figure 4. Fuzzy Lattice: MTAH

This very simple sorting procedure gives us for each many-valued attribute the distribution of the objects in the line diagram of the chosen fuzzy scale. Usually, we are interested in the interaction between two or more fuzzy many-valued attributes. This interaction can be visualized using the so-called fuzzy nested line diagrams. It is used for visualizing larger fuzzy concept lattices, and combining fuzzy conceptual scales on-line. Figure 4 shows the fuzzy nested lattice constructed from Figure 1 and 2.

### 5.4 Fuzzy Ontology Generation step

This step consists on **construct on Fuzzy Ontology**. It consists to deduce the Fuzzy Cluster Lattice corresponding to MTAH lattice generated in the first step, then to generate **Ontology Extent and Intent Classes, Ontology hierarchical Classes**. **Ontology Relational Classes and Fuzzy Ontology.**

### 5.4.1 FCL Generation

The goal of this phase is make a certain abstraction on the list of the objects with their degrees of membership in the clusters. This lattice will use to build a core of ontology.

**Definition.** A *fuzzy Clusters Lattice* (FCL) of a Fuzzy Formal Concept Lattice, is consist on a Fuzzy concept lattice such as each equivalence class (i.e. a node of the lattice) contains only the intentional description (intent) of the associated fuzzy formal concept.

We make in this case a certain abstraction on the list of the objects with their degrees of membership in the clusters. The nodes of FCL are clusters ordered by the inclusion relation. As shown from the Figure 5, we obtain a lattice more reduced, simpler to traverse and stored.

### 5.4.2 Hierarchical Relation Generation.

This step consists on Extraction of fuzzy ontology. For this, we must construct in the first, a concept hierarchy from the conceptual clusters, we need to find the hierarchy relations from the clusters. We first define a concept hierarchy as follows:

**Definition (Concept Hierarchy).** A concept hierarchy is a poset (partially ordered set) (H,<), where H is a finite set of concepts and < is a partial order on H.

Figure 6, illustrates the hierarchical relations constructed from the conceptual clusters given in Figure 5. Each concept in the concept hierarchy is represented by a set of its attributes.

The supremum and infimum of the lattice are considered as "Thing" and "Nothing" concepts, respectively.

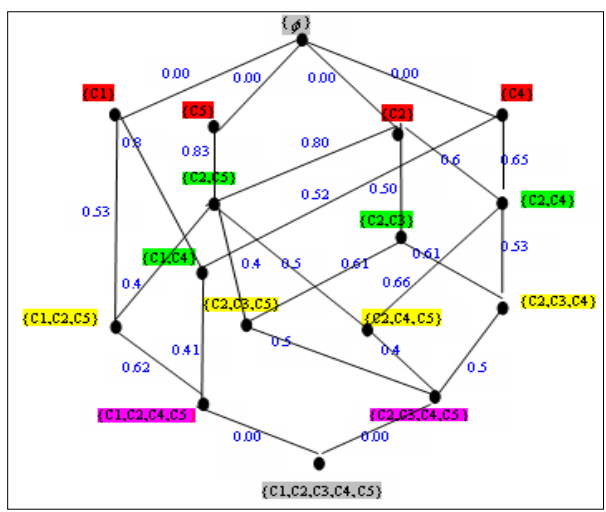

Figure 5. Fuzzy Clusters Lattice FCL

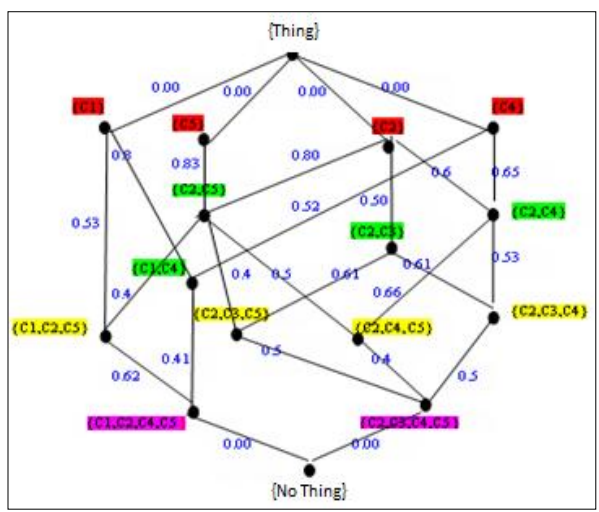

Figure 6. Fuzzy Ontology Lattice

### 5.4.3 Fuzzy Ontology Generation

This step constructs fuzzy ontology from a fuzzy context using the concept hierarchy created by fuzzy conceptual clustering. This is done based on the characteristic that both FCA and ontology support formal definitions of concepts. Thus, we define the fuzzy ontology as follows:

**Definition (Fuzzy Ontology).** A fuzzy ontology $F_o$ consists of four elements $(C, A^C, R, X)$, where :

- C represents a set of concepts,
- $A^C$ represents a collection of attributes sets, one for each concept,
- $R = (R_T; R_N)$ represents a set of relationships, which consists of two elements: $R_N$ is a set of nontaxonomy relationships and $R_T$ is a set of taxonomy relationships.
- Each concept $c_i$ in C represents a set of objects, or instances, of the same kind.
- Each object $o_{ij}$ of a concept $c_i$ can be described by a set of attributes values denoted by $A^C(c_i)$.
- Each relationship $r_i(c_p, c_q, \alpha)$ in R represents a *fuzzy association* between concepts $c_p$ and $c_q$, and the instances of such a relationship are pairs of

- ($c_p$, $c_q$) concept objects with confidence α; α is in ]0..1].
- Each attribute value of an object or relationship instance is associated with a fuzzy membership value between [0,1] implying the uncertainty degree of this attribute value or relationship.
- X is a set of axioms. Each axiom in X is a constraint on the concept's and relationship's attribute values or a constraint on the relationships between concept objects

In our approach, we consider the *Fuzzy Ontology Lattice* as a formal domain-specific ontology. This ontology has all lattice properties, which are useful for ontology sharing, reasoning about concepts as well as navigating and retrieving of information.

The whole process to create a fuzzy ontology was completely automatic. We may consider nodes as concepts. The name of the concept is a concatenation of an attribute and its label linguistics, in accordance with the correspondence in Table 4. Nevertheless, taxonomic relationships between concepts are present in the lattice.

**Example.**
C1 is transformed in Salary(Low); C2 is transformed in Salary(Medium); C3 is transformed in Salary(High); C4 is transformed in Age(Young) and C5 is transformed in Age(Adult)
r (C5 , C2, 0.83) is transformed in
r(Age(Adult), Salary(Medium),0. 83)

### 5.4.4 Semantic Representation Conversion

The generated fuzzy ontology provides a conceptual model of knowledge in the corresponding domain. However, in order to make such knowledge accessible and sharable on the Web environment, we must convert it into a semantic representation that can be embedded into the contents of Web pages. In Semantic Web, ontology description language such as fuzzy-OWL2 can be used to annotate ontology. Therefore, the generated fuzzy ontology can be automatically converted into the corresponding semantic representation in fuzzy-OWL2, in which each class and instance is annotated.

## 6 ADVANTAGES AND VALIDATION OF THE PROPOSED APPROACH

### 6.1 Advantages of the Proposed Approach

No days, there exist a few proposals for ontologies of data mining using FCA, but all of them way the starting data unit, after having done a data cleansing step and an elimination of invalid-value elements. In our point of view, the limits of these approaches consist in extracting this ontology departing from the data or a data variety, which may be huge.

In our approach, the generation of the ontology takes in consideration another degree of granularity into the process of this generation. Indeed, we propose to define ontology between classes resulting from a preliminary classification on the data. The data classification is to divide a data set into subsets, called classes, so that all data in the same class are similar and data from different classes are dissimilar. Thus:

- The number of clusters generated by a classification algorithm is always less than the number of objects starting on which we apply the classification algorithm.
- All objects belonging to the same cluster have the same properties.

This idea is in our opinion very important, view on the data set which is very voluminous. This models a certain abstraction of the data that is fundamental in the case of an enormous number of data. In this case, we define our ontology between the clusters.

### 6.2 Validation of the Proposed Approach

To validate our approach, we used Protégé 4.2[1],that support the fuzzy concept, to model our ontology described with the lattice presented in Figure 6. The figure 7 presents an excerpt of this ontology. Using Protégé 4.2, we have succeeds to generate automatically the description of our ontology with fuzzy-OWL 2 language.

As an application, we were able to evaluate queries using the appropriate interface provided by protected 4.2. This is illustrate by Figure 8.

---

[1] http://protege.stanford.edu/download/protege/4.2/

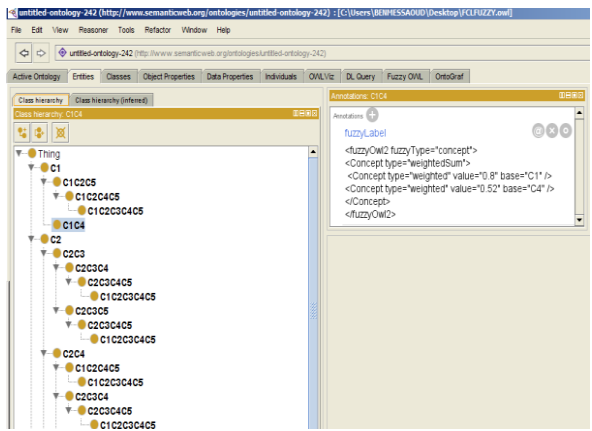

Figure 7. Fuzzy Ontology described by FuzzyOWL2 using Protected 4.2

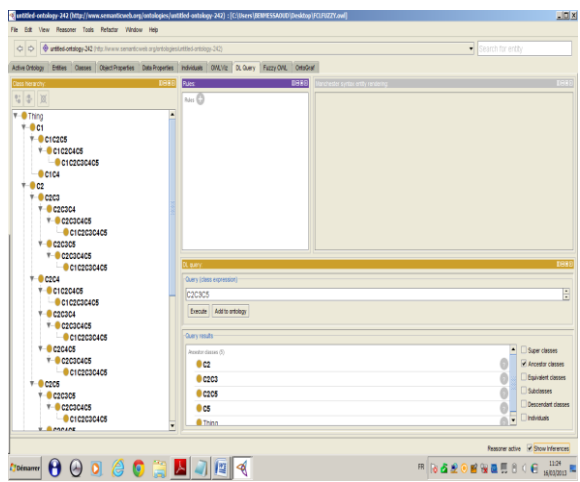

Figure 8. Fuzzy Example of evaluate query with Fuzzy Ontologie

## 7 CONCLUSION

In this paper, our challenge is to combine Clustering, FCA and Ontology in order to improve it. For this, we propose a new approach for automatic generation of Fuzzy Ontology of Data Mining, called *FODM*.

The FODM approach starts by the organization of the data in homogeneous clusters having common properties which permits to deduce the data's semantic. Then, it models these clusters by an extension of the FCA, called Fuzzy Cluster Lattice. This lattice will be used to build a core of ontology. Finally, the generate fuzzy ontology is represented using Fuzzy OWL2. To validate our approach, we used Protégé 4.2,that support the fuzzy concept, to model our ontology and to generate the script in fuzzy-OWL 2 language.

Knowing that the number of classes is always lower than the number of starting data, we prove that this solution reduced considerably the definition of the ontology, offered a better interpretation of the data and optimized both the space memory and the execution time.

As future perspectives of this work, we mention to test our approach on several large data sets.